\documentclass{article}
\usepackage[utf8]{inputenc}
\usepackage{graphics}
\usepackage{hyperref}
\usepackage[margin=0.8in]{geometry}
\usepackage{todonotes}
\usepackage{array}
\usepackage{algorithmic}
\usepackage{tabularx}
\usepackage{amsmath}
\usepackage{braket}
\newcolumntype{C}[1]{>{\centering}m{#1}}
\newcolumntype{R}[1]{>{\raggedleft}m{#1}}
\usepackage{float}
\usepackage{enumitem}
\usepackage{authblk}
\usepackage[sorting = none, backend = bibtex, style=numeric-comp]{biblatex}

\title{Analysis of a Programmable Quantum Annealer as a Random Number Generator}

\author[1]{Elijah Pelofske\thanks{Email: epelofske@lanl.gov}}
\affil[1]{Los Alamos National Laboratory, CCS-3 Information Sciences}

\begin{document}
\date{\vspace{-7ex}}

\maketitle

\begin{abstract}
Quantum devices offer a highly useful function - that is generating random numbers in a non-deterministic way since the measurement of a quantum state is not deterministic. This means that quantum devices can be constructed that generate qubits in a uniform superposition and then measure the state of those qubits. If the preparation of the qubits in a uniform superposition is unbiased, then quantum computers can be used to create high entropy, secure random numbers. Typically, preparing and measuring such quantum systems requires more time compared to classical pseudo random number generators (PRNGs) which are inherently deterministic algorithms. Therefore, the typical use of quantum random number generators (QRNGs) is to provide high entropy secure seeds for PRNGs. Quantum annealing (QA) is a type of analog quantum computation that is a relaxed form of adiabatic quantum computation and uses quantum fluctuations in order to search for ground state solutions of a programmable Ising model. Here we present extensive experimental random number results from a D-Wave 2000Q quantum annealer, totaling over 20 billion bits of QA measurements, which is significantly larger than previous D-Wave QA random number generator studies. Current quantum annealers are susceptible to noise from environmental sources and calibration errors, and are not in general unbiased samplers. Therefore, it is of interest to quantify whether noisy quantum annealers can effectively function as an unbiased QRNG. The amount of data that was collected from the quantum annealer allows a comprehensive analysis of the random bits to be performed using the NIST SP 800-22 Rev 1a testsuite, as well as min-entropy estimates from NIST SP 800-90B. The randomness tests show that the generated random bits from the D-Wave 2000Q are biased, and not unpredictable random bit sequences. With no server-side sampling post-processing, the $1$ microsecond annealing time measurements had a min-entropy of $0.824$. 
\end{abstract}


\section{Introduction}
\label{section:introduction}

Random number generation (RNG) is a very important capability in information computing. In particular \emph{unbiased} random number generation is extremely important in many computing applications. Pseudo-Random Number Generators (PRNGs) are deterministic and very fast software level algorithms that can reliably generate random numbers. True Random Number Generators (TRNGs) are based on a physical property of a system that makes the random number generation inherently non-deterministic. Quantum systems have this property of non-determinism where it is not possible to know deterministically what the measured state of a quantum system will be before it has been measured. 

Testing for randomness, in particular secure and unbiased randomness, is not directly possible. Instead, tests for patterns and biases that are clearly \emph{not} random can be tested for \cite{rukhin2001statistical, 9069949, 10155271, 8396276, electronics12030723}. If a proposed RNG is tested against enough of these tests which can detect non-random data, then you can be reasonably confident in the ability of the RNG to generate uniformly random numbers. 

One of the types of programmable quantum computers that have become available to test, typically as cloud computing resources, are D-Wave quantum annealers. Quantum annealing is a specialized type of quantum computation that aims to sample the optimal solution(s) of a combinatorial optimization problem, ideally using adiabatic evolution \cite{Finnila_1994, santoro2006optimization, Kadowaki_1998, morita2008mathematical, das2008colloquium}. Quantum annealing hardware is typically implemented using the transverse driving Hamiltonian where the system is initialized in the groundstate of the Transverse-field Hamiltonian \cite{hauke2020perspectives, Kadowaki_1998, morita2008mathematical, das2008colloquium, santoro2006optimization, santoro2002theory}. D-Wave quantum annealers are physically implemented using programmable superconducting flux qubits \cite{johnson2011quantum, boixo2013experimental, PhysRevX.4.021041, Venturelli_2015, harris2018phase, boixo2016computational, king2021scaling}. Quantum annealers, and more generally quantum computers, are potentially interesting as secure entropy sources for generating random numbers because of the inherent stochasticity of measuring quantum states - there is not a deterministic mechanism to compute what the measured state will be of an arbitrary quantum state. For this reason, quantum computers, and more generally physical sources of measurements of quantum information, are True Random Number Generators (TRNGs) (or QRNGs) \cite{RevModPhys.89.015004, ma2016quantum, 8396276}. Importantly, there exist current technologies which are secure, high bit-rate, QRNGs \cite{Bai_2021, Guo:19}. 

The primary reason that modern quantum annealers are not perfect random number generators is because there are a large number of sources of error and bias in the computation, for example the spin bath polarization effect \cite{lanting2020probing, PRXQuantum.1.020320} can cause sequential anneal-readout cycles to have self correlations (in time), and programmed coefficients (even if they are $0$) have slightly different effective weights on the hardware \cite{https://doi.org/10.48550/arxiv.1502.02098}. Furthermore, it has been shown that modern D-Wave quantum annealers have a measurable performance change over time \cite{https://doi.org/10.48550/arxiv.2209.05648, https://doi.org/10.48550/arxiv.2006.13440}. There have also been cross-qubit correlations observed on a D-Wave 2000Q chip \cite{park2023spatial, krauss2021statistical}. There have been studies which aim to reduce biases and noise present in minor-embedded QA computations, which in the case of reducing biases in the constraint of the graph partitioning problem results in effectively attempting to create an unbiased Quantum Annealing random number sampler, see ref. \cite{10.1145/3457388.3458672}. Interestingly, quantum annealing (even in an ideal computation, with no noise), does not in general sample degenerate groundstates (i.e. optimal solutions of the combinatorial optimization problem) uniformly due to the transverse field driving Hamiltonian \cite{matsuda2009quantum, PhysRevA.100.030303, PhysRevLett.118.070502, PhysRevE.99.063314, zhang2017advantages}, meaning that using the QA sampling of the groundstates of non-trivial Hamiltonians would not be a good source of unbiased random numbers. Instead, the much simpler case of an all $0$ coefficient Ising model is the most direct way to program these devices to produce random numbers (see more details in Section \ref{section:methods_QA}). D-Wave quantum annealers have been evaluated, on somewhat small problem sizes, for the possibility of utilizing them as TRNGs \footnote{Sometimes the acronym that is used for quantum devices generating random numbers is Quantum Random Number Generator (QRNG)} in previous studies \cite{9923932, LANL-DWave-QRNG}.

Quantum random number generators in general are a topic of much interest, for example there have been several studies which examined using gate model quantum computers as random number generators \cite{Tamura_2020, li2021quantum, jacak2020quantum, cryptoeprint:2020/078, 9605294, sinha2023programmable}, boson sampling \cite{shi2022unbiased}, using quantum walks to generate random numbers \cite{Sarkar_2019, yang2016novel}, and device-independent secure random number generation \cite{Liu_2018}. The idea of using random dense quantum volume circuits (see refs. \cite{PhysRevA.100.032328, Baldwin_2022, Pelofske_2022} for details on quantum volume circuits) as random number generators in the gate model setting have also been proposed \cite{e25040607}.

This paper presents the most comprehensive review of using a quantum annealer as a random number generator performed to date, totalling over 20 billion bits of qubit measurements, and testing $8$ different QA device settings for how they impact the measured bits. In particular, this very large dataset allows all of the NIST SP800-22 randomness tests, and all of the NIST SP800-90B min-entropy (non-IID) tests, to be executed on the data (some of the tests have a minimum bit length requirement). This has not been able to be done before for quantum annealing random bits \cite{9923932, LANL-DWave-QRNG} or more generally for cloud accessible quantum computers.

All data from this study are publicly available as a Zenodo dataset \cite{pelofske_2024_10583977}.

\section{Methods}
\label{section:methods}
Section \ref{section:methods_QA} details the Quantum Annealing implementation details, and Section \ref{section:methods_randomness_tests} details the randomness testsuites that are used. 

\subsection{Quantum Annealing}
\label{section:methods_QA}
The computation performed by D-Wave quantum annealers is described by eq.~\eqref{equation:QA_Hamiltonian}, and eq.~\eqref{eq:Ising} describes the discrete optimization Ising model that a user can program to be sampled by the quantum annealer (the quadratic coefficients are subject to the constraint of the native connectivity of the quantum annealing hardware). The functions $A(s)$ and $B(s)$ define the Transverse field driving Hamiltonian strength and the programmed Ising model Hamiltonian, respectively, parameterized by the variable $s$. In standard quantum annealing, which is the setting used in this study, $s$ defines a linear schedule as a function of anneal time, and the strengths of $A(s)$ and $B(s)$ at each $s$ step are system defined quantities. At the beginning of the anneal the $A(s)$ term dominates, and then over the course of the anneal $A(s)$ is reduced in strength and $B(s)$ is increased in strength.  Eq.~\eqref{eq:Ising} is a slight re-formulation of the Ising model defined in the second summation term of eq.~\eqref{equation:QA_Hamiltonian}. The goal of the quantum annealer is to find a minimum variable assignment vector $z$ given the objective function eq.~\eqref{eq:Ising}. The variable states can be either $\{0, 1\}^n$, in which case the combinatorial optimization problem is a Quadratic Unconstrained Binary Optimization (QUBO) problem, or the variables can be spins $\{+1, -1\}^n$, in which case the combinatorial optimization problem is an Ising model.

\begin{equation}
    H = - \frac{A(s)}{2} \Big( \sum_i \hat{\sigma}_{x}^{(i)} \Big) + \frac{B(s)} {2} \Big( \sum_i h_i \hat{\sigma_z}^{(i)} + \sum_{i>j} J_{i, j} \hat{\sigma_z}^{(i)} \hat{\sigma_z}^{(j)} \Big)
    \label{equation:QA_Hamiltonian}
\end{equation}

\begin{align}
    f(z_1,\ldots,z_n) = \sum_{i=1}^n h_i z_i + \sum_{i<j} J_{ij} z_i z_j,
    \label{eq:Ising}
\end{align}

\begin{figure*}[h!]
    \centering
    \includegraphics[width=0.59\textwidth]{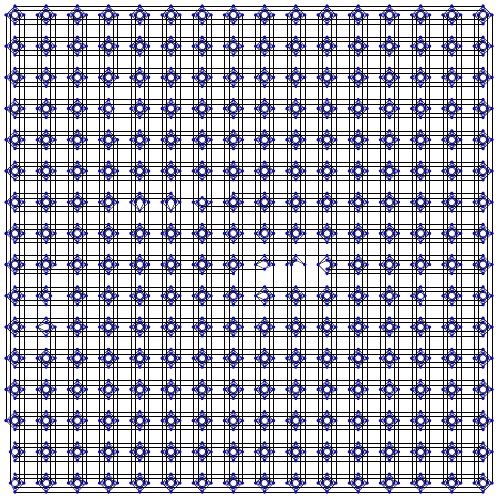}
    \caption{LANL D-Wave 2000Q hardware connectivity graph (the name of this type of connectivity graph is Chimera, which is in general a sparse but scalable hardware implementation of quantum annealing). This device has 2032 active qubits (due to hardware defects the full Chimera lattice of $2048$ qubits is not active). The chip id of this device is \texttt{DW\_2000Q\_LANL}. }
    \label{fig:LANL_Chimera}
\end{figure*}

The D-Wave quantum annealer that is used to generate random bits is \texttt{DW\_2000Q\_LANL}, the Chimera hardware graph for this device is shown in Figure \ref{fig:LANL_Chimera}. The simplest way to generate random bits using a quantum annealer is simply to set the user programmed coefficients for all linear terms (e.g. hardware qubits) and quadratic terms (e.g. hardware couplers) to $0$, meaning that only the transverse field Hamiltonian is present in the computation (ideally), which means that the qubits are in a uniform superposition, while the computation is coherent, during the anneal. There are certainly more complicated ways that could be utilized with the goal of extracting good random bits, such as random circuit sampling on gate model devices \cite{e25040607} or by tuning devices biases to improve sampling of balanced partitions for the graph partitioning problem \cite{10.1145/3457388.3458672}, however in general RNG's need to be as fast as possible and therefore minimizing the complexity of the computation is likely a good motivation to aim for. Explicitly, the Ising model (variable states are $\in \{+1, -1\}$) that we will sample is given in eq.~\eqref{eq:Ising_random}. 

\begin{align}
    f(z_1,\ldots,z_n)  = \sum_{i=1}^n 0 z_i + \sum_{i<j} 0 z_i z_j,
    \label{eq:Ising_random}
\end{align}

In many QRNG systems readout time is much longer than comparable PRNG's, and for cloud based quantum annealers this is also an important aspect to consider. In the experiments we perform, the total annealing time will be varied from $1$ microsecond up to $2000$ microseconds ($1$ microsecond is the shortest annealing time available on the D-Wave 2000Q devices). The potentially relevant thing that can be investigated is whether the $1$ microsecond annealing times can give high quality random bits, because this utilizes a relatively small amount of compute time (certainly compared to longer annealing times).

The main D-Wave parameters that will be varied are the \emph{annealing time}, which will use $1$, $10$, $100$, and $2000$ microseconds. These annealing times span the range of allowed annealing times on the \texttt{DW\_2000Q\_LANL} chip; $1$ microsecond is the smallest annealing time and $2000$ is the longest available annealing time. The other parameter that will be tested is turning on server side classical post processing which aims to improve \emph{sampling} (although in this specific case, the sampling is being done on an all zero coefficient Ising model). In order to turn on this server side post processing, the user facing post process option was set to \texttt{sampling} \cite{D_Wave_postprocessing}. The \texttt{sampling} server side post-processing option performs local bit changes on the measurements (before sending the results to the user) with the goal of obtaining a post-processes set of samples that corresponds to a Boltzmann distribution with inverse temperature $\beta$, where $\beta$ is $\beta$ is set to a value near to the inverse temperature corresponding to the raw samples (see ref. \cite{D_Wave_postprocessing} for more details). We test turning this server-side post processing option on and off. The reasoning is that we would ideally want to the quantum annealer to be able to produce unbiased random bits without this post processing, however it may be the case that the classical post processing helps reduce bias in the samples at with a small computational overhead, in which case it would be interesting to quantify this. Therefore, in total there will be $8$ datasets, each using a different quantum annealer parameter choice. Each of the datasets will be strictly sequential in time - e.g. the order of the bits will not be changed by some other entropy source. This time series representation of the data is especially important since it has been shown that there are long term trends that can be observed in current D-Wave quantum annealing processors \cite{https://doi.org/10.48550/arxiv.2209.05648}. Additionally, the exact ordering of the bits within each anneal (e.g. whole-chip readout cycle) is strictly based on the logical qubit indexing within the hardware, which is fixed for all samples, but is arbitrarily set. Each anneal-readout cycle is concatenated with the next anneal-readout cycle that was executed in time - no other source of entropy is present in the data. 

With the goal of mitigating the spin bath polarization effect \cite{lanting2020probing} self-sample correlations, all of the data is constructed by sequentially calling the D-Wave backend for a single anneal-readout cycle (i.e. instead of measuring many anneals in a single job). Each job is sampled as an Ising model, meaning that spins are the measured states (although whether the model was specified as QUBO or Ising would in principle have no impact on the results). Additionally, although the bits are all time ordered, because of network interruptions or device power losses there are gaps in the time of the sequential random bit sampling. 

The parameters used for the $8$ datasets is as follows:
\begin{enumerate}[noitemsep]
    \item \textit{Test 1} uses server side classical post-processing, and an annealing time of $1$ microseconds. 
    \item \textit{Test 2} uses server side classical post-processing, and an annealing time of $2000$ microseconds. 
    \item \textit{Test 3} uses server side classical post-processing, and an annealing time of $10$ microseconds. 
    \item \textit{Test 4} uses server side classical post-processing, and an annealing time of $100$ microseconds. 
    \item \textit{Test 5} uses default sampling with no server side classical post-processing, and an annealing time of $1$ microseconds. 
    \item \textit{Test 6} uses default sampling with no server side classical post-processing, and an annealing time of $2000$ microseconds. 
    \item \textit{Test 7} uses default sampling with no server side classical post-processing, and an annealing time of $10$ microseconds. 
    \item \textit{Test 8} uses default sampling with no server side classical post-processing, and an annealing time of $100$ microseconds. 
\end{enumerate}

For all $8$ datasets, the~\texttt{programming\_thermalization} and \texttt{readout\_thermalization} are set to $0$ microseconds so as to remove any thermalization effects (beyond thermalization that occurs after the qubits lose coherence; the qubit coherence times are estimated to be on the order of 10's of nanoseconds \cite{King_2022}). All other parameters are set to default. The motivation for evaluating these different parameter choices is the following. 

\begin{enumerate}[noitemsep]
    \item Although in general increasing annealing times on D-Wave quantum annealers results in better sampling success rate of combinatorial optimization problems, these long annealing times are much longer than the qubit coherence times of current D-Wave quantum annealers \cite{King_2023_5000q, King_2022}. Therefore, the longer annealing times are using thermalization to marginally improve the sampling success probability \cite{dickson2013thermally}. However, in this case there is not a combinatorial optimization problem being sampled - therefore in principle it may be the case that longer anneal times accumulate more errors in the computation, in particular more biases in the random sampling computation we aim to perform. Therefore, it may be advantageous to sample using the shortest annealing times available on the hardware -- whether this is true or not is the aim of the varying annealing times. Furthermore, sampling rate for random numbers is extremely important -- faster random bit sampling is more useful than slower sampling. Therefore, if the smaller annealing times produce high entropy random bits this would be better than using longer annealing times. 
    \item The server side classical post-processing is not ideal since we wish to evaluate whether the bare quantum annealing hardware can produce good random bits. However, this post processing is intended to improve sampling of combinatorial optimization problems, and in this case there is nothing to optimize with respect to energy. Nevertheless, it is an interesting question to consider whether there is a clear difference between the QA sampling with and without the server side post processing for sampling optimization. 
\end{enumerate}

\subsection{Testing for randomness}
\label{section:methods_randomness_tests}

The randomness test that will be applied to the data are all of the tests from SP800-22 Rev 1a by National Institute of Standards and Technology \cite{8966}, titled \textit{A Statistical Test Suite for Random and Pseudorandom Number Generators for Cryptographic Applications}. This testsuite contains 15 randomness tests, two of which contain several sub-tests. In total each of the $8$ datasets will be tested against $38$ randomness tests, each giving a p-value output. For the purposes of maintaining consistency, and using the original NIST SP 800-22 test definitions \cite{8966}, a computed P-value which is $\geq 0.01$ would accept the sequence as being random, and otherwise we would consider it to be non-random. This p-value threshold criteria is applied to all of the randomness tests. In the tests where there are multiple computed P-values, such as \texttt{cumulative sums} where there is a forward and backward mode of operation, all P-values must be greater than or equal to $0.01$ to be considered random. The \texttt{serial} test also outputs two P-values.

The implementation used for this analysis is the Python 3 package nistrng \footnote{\url{https://github.com/InsaneMonster/NistRng}}; this package was chosen primarily for its compatibility with NumPy \cite{arris2020array} arrays, which was necessary for the size of the datasets being tested. Other implementations that are, for example, based on casting the bits to integers does not scale well to these large dataset sizes.

The randomness test implementation details and references are not enumerated here - all details can be found in ref. \cite{8966} along with the linked open source code implementations. 

In the context of verifying entropy sources, a useful measure is the \emph{min-entropy}, which is a conservative measure of entropy sources. The \emph{min-entropy} metric gives a clear way of determining how unpredictable a set of random variable samples is, and therefore is another way of quantifying bitstring randomness. Min-entropy is maximized for a uniform distribution, as with standard Shannon entropy \cite{6773024}, and minimized closer to $0$ for biased distributions. In this case, we apply the NIST SP 800-90B testsuite (titled \emph{Recommendation for the Entropy Sources Used for Random Bit Generation}) \cite{turan2018recommendation} \footnote{\url{https://github.com/usnistgov/SP800-90B_EntropyAssessment}} in order to compute \emph{min-entropy} estimates on the quantum annealing bitstrings. In particular, the \emph{non IID} testsuite is executed in order to obtain the $h_\text{Original}$ \texttt{min-entropy} estimates from the $10$ tests in the testsuite. The NIST SP 800-90B non-IID track is intended to be applied to noise sources that do not generate Independent and Identically Distributed (IID) samples. The testsuite was executed with the help of Charliecloud containerization \cite{10.1145/3126908.3126925}.

\begin{table*}[th!]
\begin{center}
\begin{tabular}{ |p{1.9cm}||p{1.6cm}|p{1.6cm}|p{1.6cm}|p{1.6cm}|p{1.6cm}|p{1.6cm}|p{1.6cm}|p{1.6cm}| }
 \hline
 Parameter combination & Test 1 & Test 2 & Test 3 & Test 4 & Test 5 & Test 6 & Test 7 & Test 8 \\ 
 \hline
 \hline
 Bits & 2563745952 & 2573540192 & 2540168656 & 2553250672 & 2610469760 & 2580696896 & 2574846768 & 2580371776 \\ 
 \hline
\end{tabular}
\end{center}
\caption{Each dataset with different D-Wave settings contains at least $2.5$ billion time ordered bits. }
\label{table:bit_counts}
\end{table*}

\begin{table*}[th!]
\begin{center}
\begin{tabular}{ |p{5.1cm}||p{1.2cm}|p{1.2cm}|p{1.2cm}|p{1.2cm}|p{1.1cm}|p{1.1cm}|p{1.1cm}|p{1.1cm}| } 
 \hline
 Test name & Test 1 & Test 2 & Test 3 & Test 4 & Test 5 & Test 6 & Test 7 & Test 8 \\ 
 \hline
 \hline
 Monobit & $0.00298$ & $0.2144$ & $0.03651$ & $0.26875$ & $0.0$ & $0.0$ & $0.0$ & $0.0$ \\ 
 \hline
 frequency within block & $0.26277$ & $0.81603$ & $0.31771$ & $0.44942$ & $0.0$ & $0$ & $0.0$ & $0.0$ \\ 
 \hline
 Runs & $0.2031$ & $0.81369$ & $0.07247$ & $0.06147$ & $0.0$ & $0.0$ & $0.0$ & $0.0$ \\ 
 \hline
 Longest runs in a block & $0.42578$ & $0.05784$ & $0.42767$ & $0.37912$ & $0.0$ & $0.00002$ & $0.0$ & $0.00002$ \\ 
 \hline
 binary matrix rank & $0.41732$ & $0.13034$ & $0.87238$ & $0.23248$ & $0.8351$ & $0.6680$ & $0.27403$ & $0.42267$ \\ 
 \hline
 Spectral (dft) & $0.5188$ & $0.97025$ & $0.11017$ & $0.21241$ & $0.0$ & $0.0$ & $0$ & $0.70141$ \\ 
 \hline
 non overlapping template matching & $1.0$ & $1.0$ & $1.0$ & $1.0$ & $1.0$ & $1.0$ & $0.99999$ & $1.0$ \\ 
 \hline
 overlapping template matching & $0.0$ & $0.0$ & $0.0$ & $0.0$ & $0.0$ & $0.0$ & $0.0$ & $0.0$ \\ 
 \hline
 maurers universal & $0.87378$ & $0.41008$ & $0.32131$ & $0.90424$ & $0.0$ & $0.0$ & $0.0$ & $0.0$ \\ 
 \hline
 linear complexity & $0.02116$ & $0.90444$ & $0.66187$ & $0.40974$ & $0.07821$ & $0.72340$ & $0.8196$ & $0.96489$ \\ 
 \hline
 Serial & $(0.15763, \newline 0.97128)$ & $(0.48585, \newline 0.36874)$ & $(0.08526, \newline 0.4088)$ & $(0.08526, \newline 0.40881)$ & ($0$, $0$) & ($0$, $0$) & ($0$, $0$) & $(0, 0)$ \\ 
 \hline
 Approximate entropy & $0.15762$ & $0.48584$ & $0.04653$ & $0.08525$ & $0.0$ & $0$ & $0.0$ & $0.0$ \\ 
 \hline
 cumulative sums & $(0.00463, \newline 0.00502)$ & $(0.31615, \newline 0.23325)$ & $(0.05785, \newline 0.06310)$ & $(0.33323, \newline 0.46560)$ & ($0$, $0$) & ($0$, $0$) & ($0$, $0$) & ($0$, $0$) \\ 
 \hline
 random excursion $x = -4$ & $0.81758$ & $0.77005$ & $0.89501$ & $0.01337$ & $0.68502$ & $0.86836$ & $0.02697$ & $0.69493$ \\ 
 \hline
 random excursion $x = -3$ & $0.42183$ & $0.50305$ & $0.95679$ & $0.28226$ & $0.44370$ & $0.76127$ & $0.04336$ & $0.83775$ \\ 
 \hline
 random excursion $x = -2$ & $0.95911$ & $0.61655$ & $0.92114$ & $0.0906$ & $0.39231$ & $0.50231$ & $0.000001$ & $0.71308$ \\ 
 \hline
 random excursion $x = -1$ & $0.92394$ & $0.06198$ & $0.44488$ & $0.21993$ & $0.05059$ & $0.39755$ & $0.01036$ & $0.68146$ \\ 
 \hline
 random excursion $x = 1$ & $0.64727$ & $0.07625$ & $0.19236$ & $0.16096$ & $0.35359$ & $0.25673$ & $0.96257$ & $0.4132$ \\ 
 \hline
 random excursion $x = 2$ & $0.24511$ & $0.58911$ & $0.11843$ & $0.45651$ & $0.33907$ & $0.01470$ & $0.99698$ & $0.88766$ \\ 
 \hline
 random excursion $x = 3$ & $0.17254$ & $0.88776$ & $0.13784$ & $0.24313$ & $0.37950$ & $0.00178$ & $0.99922$ & $0.11974$ \\ 
 \hline
 random excursion $x = 4$ & $0.7213$ & $0.27047$ & $0.5715$ & $0.41976$ & $0.59339$ & $0.00011$ & $0.99978$ & $0.79418$ \\ 
 \hline
 random excursion variant $x = -9$ & $0.33558$ & $0.90717$ & $0.33129$ & $0.05356$ & $0.40513$ & $0.53635$ & $0.17007$ & $0.13698$ \\ 
 \hline
 random excursion variant $x = -8$ & $0.52838$ & $0.91869$ & $0.25061$ & $0.1335$ & $0.38889$ & $0.51036$ & $0.20124$ & $0.05437$ \\ 
 \hline
 random excursion variant $x = -7$ & $0.72358$ & $0.79963$ & $0.31722$ & $0.29155$ & $0.45903$ & $0.47950$ & $0.16981$ & $0.06413$ \\ 
 \hline
 random excursion variant $x = -6$ & $0.57244$ & $0.93139$ & $0.45361$ & $0.21843$ & $0.60483$ & $0.44207$ & $0.05501$ & $0.19806$ \\ 
 \hline
 random excursion variant $x = -5$ & $0.4003$ & $0.92934$ & $0.43297$ & $0.2066$ & $0.97465$ & $0.39542$ & $0.00952$ & $0.33844$ \\ 
 \hline
 random excursion variant $x = -4$ & $0.37782$ & $0.84358$ & $0.32641$ & $0.25193$ & $0.49353$ & $0.33523$ & $0.01616$ & $0.31856$ \\ 
 \hline
 random excursion variant $x = -3$ & $0.65639$ & $0.45278$ & $0.34576$ & $0.05905$ & $0.52243$ & $0.25421$ & $0.2059$ & $0.42467$ \\ 
 \hline
 random excursion variant $x = -2$ & $0.97501$ & $0.23939$ & $0.485$ & $0.01502$ & $0.69999$ & $0.21295$ & $0.22067$ & $0.68673$ \\ 
 \hline
 random excursion variant $x = -1$ & $0.94952$ & $0.89322$ & $0.48452$ & $0.0321$ & $0.50450$ & $0.11666$ & $0.1573$ & $0.43766$ \\ 
 \hline
 random excursion variant $x = 1$ & $0.59362$ & $0.56961$ & $1.0$ & $0.03362$ & $0.07005$ & $0.07756$ & $0.4795$ & $0.81588$ \\ 
 \hline
 random excursion variant $x = 2$ & $0.97917$ & $0.66431$ & $0.9082$ & $0.04791$ & $0.07815$ & $0.07004$ & $0.68309$ & $0.75377$ \\ 
 \hline
 random excursion variant $x = 3$ & $0.41162$ & $0.6034$ & $0.88949$ & $0.1526$ & $0.23251$ & $0.21949$ & $0.75183$ & $0.29773$ \\ 
 \hline
 random excursion variant $x = 4$ & $0.20966$ & $0.44556$ & $0.48631$ & $0.57272$ & $0.54017$ & $0.94091$ & $0.78927$ & $0.19678$ \\ 
 \hline
 random excursion variant $x = 5$ & $0.12418$ & $0.22361$ & $0.51268$ & $0.84155$ & $0.56727$ & $0.89598$ & $0.81366$ & $0.48484$ \\ 
 \hline
 random excursion variant $x = 6$ & $0.03772$ & $0.13563$ & $0.74301$ & $0.65679$ & $0.47237$ & $0.51541$ & $0.83117$ & $0.52748$ \\ 
 \hline
 random excursion variant $x = 7$ & $0.02872$ & $0.14485$ & $0.93867$ & $0.61234$ & $0.52569$ & $0.30139$ & $0.84452$ & $0.30148$ \\ 
 \hline
 random excursion variant $x = 8$ & $0.04955$ & $0.28153$ & $0.77011$ & $0.64054$ & $0.47527$ & $0.41783$ & $0.85513$ & $0.3779$ \\ 
 \hline
 random excursion variant $x = 9$ & $0.06867$ & $0.38781$ & $0.79821$ & $0.70145$ & $0.39221$ & $0.3919$ & $0.86383$ & $0.54692$ \\ 
 \hline
\end{tabular}
\end{center}
\caption{Randomness test computed p-values for all of the tests within the NIST SP800-22 Rev 1a testsuite, for all $8$ parameter variations of the quantum annealing experiments, executed on \texttt{DW\_2000Q\_LANL}.  }
\label{table:NIST_randomness_tests}
\end{table*}

\section{Results}
\label{section:results}

Table \ref{table:NIST_randomness_tests} shows the complete randomness testsuite data for the $8$ QA implementation variations. Table \ref{table:bit_counts} shows the total size of each of the $8$ datasets. The threshold for failing each randomness test varies depending on the test, but a p-value less than $0.01$ definitely shows that the dataset fails that randomness test. The result is that there is no QA device setting that generates random bit strings that pass all of the randomness tests. 

Notably, the server side classical post processing did improve the random bitstring - in the sense that more of the tests passed when that post processing was applied. Also very notable is that the raw non post-processed QA data failed the monobit test, arguably the most fundamental randomness test that can be applied. This shows clearly that there was too much bias in the computation on the D-Wave 2000Q device to produce high quality random bit sequences. 

\begin{figure*}[h!]
    \centering
    \includegraphics[width=0.99\textwidth]{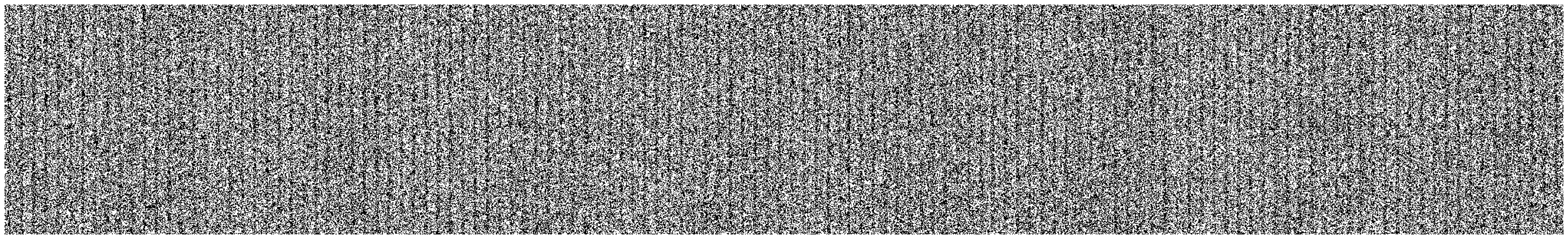}
    \caption{Bit plot of a subset of the D-Wave QRNG measurements visually showing $+1$ and $-1$ qubit states, for the $1$ microsecond annealing time and no server side post processing runs (Test 5). There are $2032$ row indices, corresponding to the $2032$ qubit indices, and $300$ column indices corresponding to $300$ time ordered anneal-readout cycles. Noticeably, there are clearly correlations in the time ordered bit sequences. }
    \label{fig:bit_sequences_5}
\end{figure*}

\begin{figure*}[h!]
    \centering
    \includegraphics[width=0.99\textwidth]{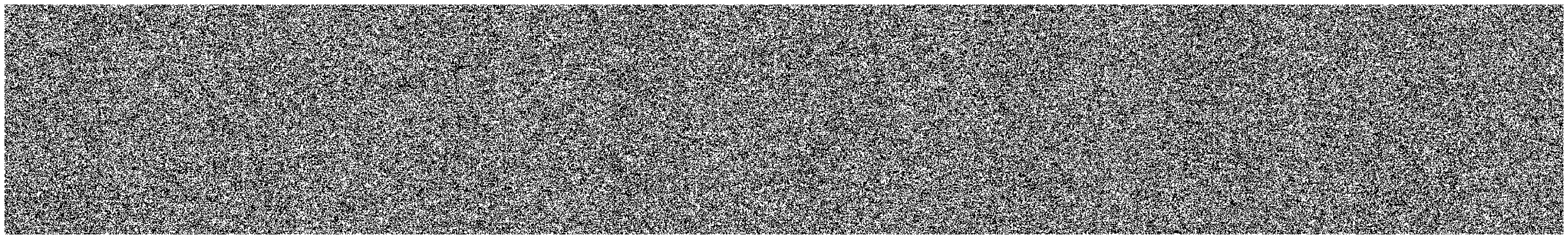}
    \caption{Bit plot of a subset of the D-Wave QRNG measurements visually showing $+1$ and $-1$ qubit states, for the $2000$ microsecond annealing time and no server side post processing runs (Test 6). There are $2032$ row indices, corresponding to the $2032$ qubit indices, and $300$ column indices corresponding to $300$ time ordered anneal-readout cycles. }
    \label{fig:bit_sequences_6}
\end{figure*}

Figures \ref{fig:bit_sequences_5} and \ref{fig:bit_sequences_6} show bit map plots of a small subset of the QA measurements, for tests 5 and 6 respectively. Notice that in Figure \ref{fig:bit_sequences_5} there are clearly time correlated trends (seen as vertical lines). Figure \ref{fig:bit_sequences_6} contains some periodic visual trends across groups of many qubits that are time correlated, but are less pronounced than Figure \ref{fig:bit_sequences_5}.

Lastly, Table \ref{table:NIST_min_entropy_tests} enumerates the \texttt{min-entropy} estimates of the entirety of the QA bitstrings for all $8$ device settings, which shows that the post-processed datasets have a higher (better) \texttt{min-entropy} compared to the raw measurements.

\begin{table*}[th!]
\begin{center}
\begin{tabular}{ |p{3.1cm}||p{1.3cm}|p{1.3cm}|p{1.3cm}|p{1.3cm}|p{1.3cm}|p{1.3cm}|p{1.3cm}|p{1.3cm}| } 
 \hline
 Test name & Test 1 & Test 2 & Test 3 & Test 4 & Test 5 & Test 6 & Test 7 & Test 8 \\ 
 \hline
 \hline
 Most Common Value & $0.999842$ & $0.999891$ & $0.999866$ & $0.999895$ & $0.923050$ & $0.963594$ & $0.926796$ & $0.989233$ \\ 
 \hline
 Collision Test & $0.988907$ & $0.988040$ & $0.986848$ & $0.983599$ & $0.861706$ & $0.954721$ & $0.897248$ & $1.000000$ \\ 
 \hline
 Markov Test & $0.999951$ & $0.999971$ & $0.999949$ & $0.999915$ & $0.914111$ & $0.963345$ & $0.923366$ & $0.990740$ \\ 
 \hline
 Compression Test & $0.980403$ & $0.965174$ & $0.974589$ & $0.978102$ & $0.824339$ & $0.847210$ & $0.828144$ & $0.888519$ \\ 
 \hline
 T-Tuple Test & $0.949592$ & $0.948716$ & $0.951325$ & $0.949373$ & $0.893867$ & $0.886097$ & $0.914859$ & $0.919266$ \\ 
 \hline
 LRS Test & $0.994078$ & $0.999435$ & $0.981888$ & $0.991583$ & $0.993062$ & $0.997555$ & $0.978171$ & $0.997016$ \\ 
 \hline
 Multi Most Common in Window Test & $0.999935$ & $0.999913$ & $0.999952$ & $0.999927$ & $0.923147$ & $0.967941$ & $0.926869$ & $0.991193$ \\ 
 \hline
 Lag Prediction Test & $0.999943$ & $0.999927$ & $0.999891$ & $0.999920$ & $0.970287$ & $0.978889$ & $0.979499$ & $0.983893$ \\ 
 \hline
 Multi Markov Model with Counting Test & $0.999889$ & $0.999882$ & $0.999883$ & $0.999938$ & $0.921688$ & $0.955856$ & $0.926797$ & $0.972710$ \\ 
 \hline
 LZ78Y Test & $0.999855$ & $0.999920$ & $0.999896$ & $0.999889$ & $0.923050$ & $0.963594$ & $0.926796$ & $0.989233$ \\ 
 \hline
 \hline
 \hline
 Overall \texttt{min-entropy} & $0.949592$ & $0.948716$ & $0.951325$ & $0.949373$ & $0.824339$ & $0.847210$ & $0.828144$ & $0.888519$ \\ 
 \hline
\end{tabular}
\end{center}
\caption{Min-entropy estimates ($h_\text{Original}$) for all quantum annealing experiment binary datasets, executed on \texttt{DW\_2000Q\_LANL}. For each of the $8$ datasets, the overall min-entropy is the minimum $h_\text{Original}$ computed across the suite of tests. The min-entropy, like standard information entropy, is maximized for a uniform distribution, and in this case (for bitstrings) the maximum entropy is $1$ and the closer the entropy is to $0$ corresponds to more biased samples. }
\label{table:NIST_min_entropy_tests}
\end{table*}

\section{Discussion and Conclusion}
\label{section:conclusion}

Even if QRNGs based on near term devices are fundamentally non-deterministic, noise present in the computation can still produce biased random bitstrings. This is what is observed in this D-Wave quantum annealer data. This is not unexpected, especially given the observed trends over time on multiple D-Wave quantum annealers \cite{https://doi.org/10.48550/arxiv.2209.05648}. However, it is important to note that this type of biased random sampling is very likely to occur with other Noisy Intermediate Scale Quantum (NISQ) computers \cite{Preskill_2018}. It is necessary that extensive tests, such as the ones presented in this paper, must be executed in order for such quantum devices to pass the threshold of being unbiased random bit samplers. Importantly, even large testsuites can not absolutely determine that the generator is indeed random - there are only tests which can show that a bit sequence is not random, or in other words the null hypothesis can never be proven to be true, it can only be observed to fail. Indeed, it has been shown that the NIST SP 800-22 testsuite is not sufficiently rigorous for verifying randomness \cite{9069949, 8166460, marton2015interpretation}. However, it does serve as a reasonable minimum threshold test, which in this case the D-Wave 2000Q device did not pass. 

Interestingly, within each QA dataset there are sometimes only a few tests which failed, but on the whole many of the tests were passed with p-values much greater than $0.01$. In terms of the \texttt{min-entropy} measure, all $4$ device settings which used server-side sampling post processing had a higher (e.g. better) min-entropy compared to the raw results. Across the battery of tests on the QA bitstrings with no post-processing, the $1$ microsecond annealing time bitstrings had a min-entropy of $0.824339$, the $2000$ microsecond annealing time bitstrings had a min-entropy of $0.847210$, the $10$ microsecond annealing time bitstrings had a min-entropy of $0.828144$, and the $100$ microsecond annealing time bitstrings had a min-entropy of $0.888519$. 

Evaluating this source of random numbers using more comprehensive testsuites, such as \emph{dieharder} \cite{brown2018dieharder} would be good - the limitation is that those tools require a significant amount of data to be analyzed (more than used in this analysis), which is currently not feasible to obtain using cloud based quantum computer access. In general, we expect that longer coherence times \cite{King_2023_5000q, King_2022} and lower error rates of manufactured quantum annealers would correspond to being able to produce higher quality random bit strings. 

A potentially interesting analysis on this existing data from \texttt{DW\_2000Q\_LANL} would be to determine if there are strong cross-qubit correlations on the chip. If such correlations exist, then this could indicate cross-talk errors from the control system.

\section{Acknowledgments}
\label{section:acknowledgments}
This work was supported by the U.S. Department of Energy through the Los Alamos National Laboratory. Los Alamos National Laboratory is operated by Triad National Security, LLC, for the National Nuclear Security Administration of U.S. Department of Energy (Contract No. 89233218CNA000001). Research presented in this article was supported by the NNSA's Advanced Simulation and Computing Beyond Moore's Law Program at Los Alamos National Laboratory. This research used resources provided by the Darwin testbed, including usage of Charliecloud containerization \cite{10.1145/3126908.3126925}, at Los Alamos National Laboratory (LANL) which is funded by the Computational Systems and Software Environments subprogram of LANL's Advanced Simulation and Computing program (NNSA/DOE). The research presented in this article was supported by the Laboratory Directed Research and Development program of Los Alamos National Laboratory under project numbers 20220656ER and 20190065DR. This research used resources provided by the Los Alamos National Laboratory Institutional Computing Program, which is supported by the U.S. Department of Energy National Nuclear Security Administration under Contract No.~89233218CNA000001. This work has been assigned the LANL report number LA-UR-23-23112. 


\setlength\bibitemsep{0pt}
\printbibliography

@incollection{Tamura_2020,
	doi = {10.1007/978-981-15-5191-8_6},
  
	url = {https://doi.org/10.1007\%2F978-981-15-5191-8_6},
  
	year = 2020,
	month = {oct},
  
	publisher = {Springer Singapore},
  
	pages = {17--37},
  
	author = {Kentaro Tamura and Yutaka Shikano},
  
	title = {Quantum Random Numbers Generated by a Cloud Superconducting Quantum Computer},
  
	booktitle = {International Symposium on Mathematics, Quantum Theory, and Cryptography}
}

@article{li2021quantum,
  title={Quantum random number generator using a cloud superconducting quantum computer based on source-independent protocol},
  author={Li, Yuanhao and Fei, Yangyang and Wang, Weilong and Meng, Xiangdong and Wang, Hong and Duan, Qianheng and Ma, Zhi},
  journal={Scientific Reports},
  volume={11},
  number={1},
  pages={1--11},
  year={2021},
  publisher={Nature Publishing Group}
}

@misc{LANL-DWave-QRNG,
author = {Sarah Michalak, Rick Picard},
title = {Leveraging LANL’s D-WAVE 2X for Random Number Generation},
month = {April},
year = {2017},
howpublished={\url{https://www.lanl.gov/projects/national-security-education-center/information-science-technology/dwave/assets/michalak_dwave2017.pdf}}
}

@misc{8966,
  author = {Lawrence Bassham and Andrew Rukhin and Juan Soto and James Nechvatal and Miles Smid and Stefan Leigh and M Levenson and M Vangel and Nathanael Heckert and D Banks},
  title = {A Statistical Test Suite for Random and Pseudorandom Number Generators for Cryptographic Applications},
  year = {2010},
  month = {2010-09-16},
  publisher = {Special Publication (NIST SP), National Institute of Standards and Technology, Gaithersburg, MD},
  url = {https://tsapps.nist.gov/publication/get_pdf.cfm?pub_id=906762},
  language = {en},
}

@article{johnson2011quantum,
  title={Quantum annealing with manufactured spins},
  author={Johnson, Mark W and Amin, Mohammad HS and Gildert, Suzanne and Lanting, Trevor and Hamze, Firas and Dickson, Neil and Harris, Richard and Berkley, Andrew J and Johansson, Jan and Bunyk, Paul and others},
  journal={Nature},
  volume={473},
  number={7346},
  pages={194--198},
  year={2011},
  publisher={Nature Publishing Group}
}

@misc{https://doi.org/10.48550/arxiv.2006.13440,
  doi = {10.48550/ARXIV.2006.13440},
  
  url = {https://arxiv.org/abs/2006.13440},
  
  author = {Suzuki, Takayuki and Nakazato, Hiromichi},
  
  title = {A proposal of noise suppression for quantum annealing},
  
  publisher = {arXiv},
  
  year = {2020},
  
  copyright = {arXiv.org perpetual, non-exclusive license}
}

@inproceedings{10.1145/3457388.3458672,
author = {Pelofske, Elijah and Hahn, Georg and Djidjev, Hristo N.},
title = {Reducing Quantum Annealing Biases for Solving the Graph Partitioning Problem},
year = {2021},
isbn = {9781450384049},
publisher = {Association for Computing Machinery},
address = {New York, NY, USA},
url = {https://doi.org/10.1145/3457388.3458672},
doi = {10.1145/3457388.3458672},
booktitle = {Proceedings of the 18th ACM International Conference on Computing Frontiers},
pages = {133–139},
numpages = {7},
location = {Virtual Event, Italy},
series = {CF '21}
}

@article{PRXQuantum.1.020320,
  title = {Simulating the Shastry-Sutherland Ising Model Using Quantum Annealing},
  author = {Kairys, Paul and King, Andrew D. and Ozfidan, Isil and Boothby, Kelly and Raymond, Jack and Banerjee, Arnab and Humble, Travis S.},
  journal = {PRX Quantum},
  volume = {1},
  issue = {2},
  pages = {020320},
  numpages = {11},
  year = {2020},
  month = {Dec},
  publisher = {American Physical Society},
  doi = {10.1103/PRXQuantum.1.020320},
  url = {https://link.aps.org/doi/10.1103/PRXQuantum.1.020320}
}

@ARTICLE{9923932,  author={Bhatia, Harshil and Tretschk, Edith and Theobalt, Christian and Golyanik, Vladislav},  journal={IEEE Access},   title={Generation of Truly Random Numbers on a Quantum Annealer},   year={2022},  volume={10},  number={},  pages={112832-112844},  doi={10.1109/ACCESS.2022.3215500}}

@article{Sarkar_2019,
	doi = {10.1038/s41598-019-48844-4},
	url = {https://doi.org/10.1038%2Fs41598-019-48844-4},
	year = 2019,
	month = {aug},
	publisher = {Springer Science and Business Media {LLC}},
  
	volume = {9},
  
	number = {1},
  
	author = {Anupam Sarkar and C. M. Chandrashekar},
  
	title = {Multi-bit quantum random number generation from a single qubit quantum walk},
  
	journal = {Scientific Reports}
}

@article{jacak2020quantum,
  title={Quantum random number generators with entanglement for public randomness testing},
  author={Jacak, Janusz E and Jacak, Witold A and Donderowicz, Wojciech A and Jacak, Lucjan},
  journal={Scientific Reports},
  volume={10},
  number={1},
  pages={1--9},
  year={2020},
  publisher={Nature Publishing Group},
  doi={s41598-019-56706-2}
}

@article{Liu_2018,
	doi = {10.1038/s41586-018-0559-3},
  
	url = {https://doi.org/10.1038%2Fs41586-018-0559-3},
  
	year = 2018,
	month = {sep},
  
	publisher = {Springer Science and Business Media {LLC}
},
  
	volume = {562},
  
	number = {7728},
  
	pages = {548--551},
  
	author = {Yang Liu and Qi Zhao and Ming-Han Li and Jian-Yu Guan and Yanbao Zhang and Bing Bai and Weijun Zhang and Wen-Zhao Liu and Cheng Wu and Xiao Yuan and Hao Li and W. J. Munro and Zhen Wang and Lixing You and Jun Zhang and Xiongfeng Ma and Jingyun Fan and Qiang Zhang and Jian-Wei Pan},
  
	title = {Device-independent quantum random-number generation},
  
	journal = {Nature}
}

@article{yang2016novel,
  title={Novel pseudo-random number generator based on quantum random walks},
  author={Yang, Yu-Guang and Zhao, Qian-Qian},
  journal={Scientific reports},
  volume={6},
  number={1},
  pages={1--11},
  year={2016},
  publisher={Nature Publishing Group},
  doi={10.1038/srep20362}
}

@misc{https://doi.org/10.48550/arxiv.1502.02098,
  doi = {10.48550/ARXIV.1502.02098},
  
  url = {https://arxiv.org/abs/1502.02098},
  
  author = {King, Andrew D. and Lanting, Trevor and Harris, Richard},
  
  title = {Performance of a quantum annealer on range-limited constraint satisfaction problems},
  
  publisher = {arXiv},
  
  year = {2015},
  
  copyright = {arXiv.org perpetual, non-exclusive license}
}

@article{https://doi.org/10.48550/arxiv.2209.05648,
	doi = {10.1088/2058-9565/accbe6},
  
	url = {https://doi.org/10.1088%2F2058-9565%2Faccbe6},
  
	year = 2023,
	month = {apr},
  
	publisher = {{IOP} Publishing},
  
	volume = {8},
  
	number = {3},
  
	pages = {035005},
  
	author = {Elijah Pelofske and Georg Hahn and Hristo N Djidjev},
  
	title = {Noise dynamics of quantum annealers: estimating the effective noise using idle qubits},
  
	journal = {Quantum Science and Technology}
}

@ARTICLE{9069949,

  author={Hurley-Smith, Darren and Patsakis, Constantinos and Hernandez-Castro, Julio},

  journal={IEEE Transactions on Information Forensics and Security}, 

  title={On the Unbearable Lightness of FIPS 140–2 Randomness Tests}, 

  year={2022},

  volume={17},

  number={},

  pages={3946-3958},

  doi={10.1109/TIFS.2020.2988505}}

@techreport{rukhin2001statistical,
  title={A statistical test suite for random and pseudorandom number generators for cryptographic applications},
  author={Rukhin, Andrew and Soto, Juan and Nechvatal, James and Smid, Miles and Barker, Elaine},
  year={2001},
  institution={Booz-allen and hamilton inc mclean va}
}

@article{PhysRevX.4.021041,
  title = {Entanglement in a Quantum Annealing Processor},
  author = {Lanting, T. and Przybysz, A. J. and Smirnov, A. Yu. and Spedalieri, F. M. and Amin, M. H. and Berkley, A. J. and Harris, R. and Altomare, F. and Boixo, S. and Bunyk, P. and Dickson, N. and Enderud, C. and Hilton, J. P. and Hoskinson, E. and Johnson, M. W. and Ladizinsky, E. and Ladizinsky, N. and Neufeld, R. and Oh, T. and Perminov, I. and Rich, C. and Thom, M. C. and Tolkacheva, E. and Uchaikin, S. and Wilson, A. B. and Rose, G.},
  journal = {Phys. Rev. X},
  volume = {4},
  issue = {2},
  pages = {021041},
  numpages = {14},
  year = {2014},
  month = {May},
  publisher = {American Physical Society},
  doi = {10.1103/PhysRevX.4.021041},
  url = {https://link.aps.org/doi/10.1103/PhysRevX.4.021041}
}

@article{boixo2013experimental,
  title={Experimental signature of programmable quantum annealing},
  author={Boixo, Sergio and Albash, Tameem and Spedalieri, Federico M and Chancellor, Nicholas and Lidar, Daniel A},
  journal={Nature communications},
  volume={4},
  number={1},
  pages={1--8},
  year={2013},
  publisher={Nature Publishing Group},
  doi={10.1038/ncomms3067}
}

@article{Venturelli_2015,
	doi = {10.1103/physrevx.5.031040},
  
	url = {https://doi.org/10.1103%2Fphysrevx.5.031040},
  
	year = 2015,
	month = {sep},
  
	publisher = {American Physical Society ({APS})},
  
	volume = {5},
  
	number = {3},
  
	author = {Davide Venturelli and Salvatore Mandr{\`{a}
} and Sergey Knysh and Bryan O'Gorman and Rupak Biswas and Vadim Smelyanskiy},
  
	title = {Quantum Optimization of Fully Connected Spin Glasses},
  
	journal = {Physical Review X}
}

@article{das2008colloquium,
  title={Colloquium: Quantum annealing and analog quantum computation},
  author={Das, Arnab and Chakrabarti, Bikas K},
  journal={Reviews of Modern Physics},
  volume={80},
  number={3},
  pages={1061},
  year={2008},
  publisher={APS},
  doi = {10.1103/revmodphys.80.1061}
}

@article{morita2008mathematical,
  title={Mathematical foundation of quantum annealing},
  author={Morita, Satoshi and Nishimori, Hidetoshi},
  journal={Journal of Mathematical Physics},
  volume={49},
  number={12},
  pages={125210},
  year={2008},
  publisher={American Institute of Physics},
  doi = {10.1063/1.2995837},
}

@article{Kadowaki_1998,
	doi = {10.1103/physreve.58.5355},
  
	url = {https://doi.org/10.1103%2Fphysreve.58.5355},
  
	year = 1998,
	month = {nov},
  
	publisher = {American Physical Society ({APS})},
  
	volume = {58},
  
	number = {5},
  
	pages = {5355--5363},
  
	author = {Tadashi Kadowaki and Hidetoshi Nishimori},
  
	title = {Quantum annealing in the transverse Ising model},
	journal = {Physical Review E}
}

@article{harris2018phase,
  title={Phase transitions in a programmable quantum spin glass simulator},
  author={Harris, R and Sato, Y and Berkley, AJ and Reis, M and Altomare, F and Amin, MH and Boothby, K and Bunyk, P and Deng, C and Enderud, C and others},
  journal={Science},
  volume={361},
  number={6398},
  pages={162--165},
  year={2018},
  publisher={American Association for the Advancement of Science}
}

@article{santoro2002theory,
  title={Theory of quantum annealing of an Ising spin glass},
  author={Santoro, Giuseppe E and Marton{\'a}k, Roman and Tosatti, Erio and Car, Roberto},
  journal={Science},
  volume={295},
  number={5564},
  pages={2427--2430},
  year={2002},
  publisher={American Association for the Advancement of Science},
  doi={10.1126/science.1068774}
}

@article{hauke2020perspectives,
  title={Perspectives of quantum annealing: Methods and implementations},
  author={Hauke, Philipp and Katzgraber, Helmut G and Lechner, Wolfgang and Nishimori, Hidetoshi and Oliver, William D},
  journal={Reports on Progress in Physics},
  volume={83},
  number={5},
  pages={054401},
  year={2020},
  publisher={IOP Publishing},
  doi={10.1088/1361-6633/ab85b8}
}

@article{RevModPhys.89.015004,
  title = {Quantum random number generators},
  author = {Herrero-Collantes, Miguel and Garcia-Escartin, Juan Carlos},
  journal = {Rev. Mod. Phys.},
  volume = {89},
  issue = {1},
  pages = {015004},
  numpages = {48},
  year = {2017},
  month = {Feb},
  publisher = {American Physical Society},
  doi = {10.1103/RevModPhys.89.015004},
  url = {https://link.aps.org/doi/10.1103/RevModPhys.89.015004}
}

@article{ma2016quantum,
  title={Quantum random number generation},
  author={Ma, Xiongfeng and Yuan, Xiao and Cao, Zhu and Qi, Bing and Zhang, Zhen},
  journal={npj Quantum Information},
  volume={2},
  number={1},
  pages={1--9},
  year={2016},
  publisher={Nature Publishing Group},
  doi={10.1038/npjqi.2016.21}
}

@misc{park2023spatial,
      title={Spatial correlations in the qubit properties of D-Wave 2000Q measured and simulated qubit networks}, 
      author={Jessica Park and Susan Stepney and Irene D'Amico},
      year={2023},
      eprint={2305.07385},
      archivePrefix={arXiv},
      primaryClass={quant-ph}
}

@article{e25040607,
AUTHOR = {Cenedese, Gabriele and Bondani, Maria and Rosa, Dario and Benenti, Giuliano},
TITLE = {Generation of Pseudo-Random Quantum States on Actual Quantum Processors},
JOURNAL = {Entropy},
VOLUME = {25},
YEAR = {2023},
NUMBER = {4},
ARTICLE-NUMBER = {607},
URL = {https://www.mdpi.com/1099-4300/25/4/607},
ISSN = {1099-4300},
DOI = {10.3390/e25040607}
}

@article{King_2022,
	doi = {10.1038/s41567-022-01741-6},
  
	url = {https://doi.org/10.1038%2Fs41567-022-01741-6},
  
	year = 2022,
	month = {sep},
  
	publisher = {Springer Science and Business Media {LLC}
},
  
	volume = {18},
  
	number = {11},
  
	pages = {1324--1328},
  
	author = {Andrew D. King and Sei Suzuki and Jack Raymond and Alex Zucca and Trevor Lanting and Fabio Altomare and Andrew J. Berkley and Sara Ejtemaee and Emile Hoskinson and Shuiyuan Huang and Eric Ladizinsky and Allison J. R. MacDonald and Gaelen Marsden and Travis Oh and Gabriel Poulin-Lamarre and Mauricio Reis and Chris Rich and Yuki Sato and Jed D. Whittaker and Jason Yao and Richard Harris and Daniel A. Lidar and Hidetoshi Nishimori and Mohammad H. Amin},
  
	title = {Coherent quantum annealing in a programmable 2,000{\hspace{0.167em}}qubit Ising chain},
  
	journal = {Nature Physics}
}

@article{King_2023_5000q,
	doi = {10.1038/s41586-023-05867-2},
  
	url = {https://doi.org/10.1038%2Fs41586-023-05867-2},
  
	year = 2023,
	month = {apr},
  
	publisher = {Springer Science and Business Media {LLC}
},
  
	volume = {617},
  
	number = {7959},
  
	pages = {61--66},
  
	author = {Andrew D. King and Jack Raymond and Trevor Lanting and Richard Harris and Alex Zucca and Fabio Altomare and Andrew J. Berkley and Kelly Boothby and Sara Ejtemaee and Colin Enderud and Emile Hoskinson and Shuiyuan Huang and Eric Ladizinsky and Allison J. R. MacDonald and Gaelen Marsden and Reza Molavi and Travis Oh and Gabriel Poulin-Lamarre and Mauricio Reis and Chris Rich and Yuki Sato and Nicholas Tsai and Mark Volkmann and Jed D. Whittaker and Jason Yao and Anders W. Sandvik and Mohammad H. Amin},
  
	title = {Quantum critical dynamics in a 5,000-qubit programmable spin glass},
  
	journal = {Nature}
}

@article{arris2020array,
 title         = {Array programming with {NumPy}},
 author        = {Charles R. Harris and K. Jarrod Millman and St{\'{e}}fan J.
                 van der Walt and Ralf Gommers and Pauli Virtanen and David
                 Cournapeau and Eric Wieser and Julian Taylor and Sebastian
                 Berg and Nathaniel J. Smith and Robert Kern and Matti Picus
                 and Stephan Hoyer and Marten H. van Kerkwijk and Matthew
                 Brett and Allan Haldane and Jaime Fern{\'{a}}ndez del
                 R{\'{i}}o and Mark Wiebe and Pearu Peterson and Pierre
                 G{\'{e}}rard-Marchant and Kevin Sheppard and Tyler Reddy and
                 Warren Weckesser and Hameer Abbasi and Christoph Gohlke and
                 Travis E. Oliphant},
 year          = {2020},
 month         = sep,
 journal       = {Nature},
 volume        = {585},
 number        = {7825},
 pages         = {357--362},
 doi           = {10.1038/s41586-020-2649-2},
 publisher     = {Springer Science and Business Media {LLC}},
 url           = {https://doi.org/10.1038/s41586-020-2649-2}
}

@article{boixo2016computational,
  title={Computational multiqubit tunnelling in programmable quantum annealers},
  author={Boixo, Sergio and Smelyanskiy, Vadim N and Shabani, Alireza and Isakov, Sergei V and Dykman, Mark and Denchev, Vasil S and Amin, Mohammad H and Smirnov, Anatoly Yu and Mohseni, Masoud and Neven, Hartmut},
  journal={Nature communications},
  volume={7},
  number={1},
  pages={10327},
  year={2016},
  publisher={Nature Publishing Group UK London},
  doi={10.1038/ncomms10327}
}

@article{dickson2013thermally,
  title={Thermally assisted quantum annealing of a 16-qubit problem},
  author={Dickson, Neil G and Johnson, MW and Amin, MH and Harris, R and Altomare, F and Berkley, AJ and Bunyk, P and Cai, J and Chapple, EM and Chavez, P and others},
  journal={Nature communications},
  volume={4},
  number={1},
  pages={1903},
  year={2013},
  publisher={Nature Publishing Group UK London},
  doi={10.1038/ncomms2920}
}

@article{king2021scaling,
  title={Scaling advantage over path-integral Monte Carlo in quantum simulation of geometrically frustrated magnets},
  author={King, Andrew D and Raymond, Jack and Lanting, Trevor and Isakov, Sergei V and Mohseni, Masoud and Poulin-Lamarre, Gabriel and Ejtemaee, Sara and Bernoudy, William and Ozfidan, Isil and Smirnov, Anatoly Yu and others},
  journal={Nature communications},
  volume={12},
  number={1},
  pages={1--6},
  year={2021},
  publisher={Nature Publishing Group},
  doi={10.1038/s41467-021-20901-5}
}

@ARTICLE{10155271,

  author={Chen, Dongyu and Chen, Hua and Fan, Limin and Luo, Kai},

  journal={IEEE Transactions on Information Forensics and Security}, 

  title={Error Analysis of NIST SP 800-22 Test Suite}, 

  year={2023},

  volume={18},

  number={},

  pages={3745-3759},

  doi={10.1109/TIFS.2023.3287391}}

@article{brown2018dieharder,
  title={Dieharder},
  author={Brown, Robert G and Eddelbuettel, Dirk and Bauer, David},
  journal={Duke University Physics Department Durham, NC},
  pages={27708--0305},
  year={2018}
}

@article{Preskill_2018,
	doi = {10.22331/q-2018-08-06-79},
  
	url = {https://doi.org/10.22331%2Fq-2018-08-06-79},
  
	year = 2018,
	month = {aug},
  
	publisher = {Verein zur Forderung des Open Access Publizierens in den Quantenwissenschaften},
  
	volume = {2},
  
	pages = {79},
  
	author = {John Preskill},
  
	title = {Quantum Computing in the {NISQ} era and beyond},
  
	journal = {Quantum}
}

@INPROCEEDINGS{8166460,

  author={Georgescu, Carmina and Simion, Emil and Nita, Alina-Petrescu and Toma, Antonela},

  booktitle={2017 9th International Conference on Electronics, Computers and Artificial Intelligence (ECAI)}, 

  title={A view on NIST randomness tests (In)Dependence}, 

  year={2017},

  volume={},

  number={},

  pages={1-4},

  doi={10.1109/ECAI.2017.8166460}}

@misc{cryptoeprint:2020/078,
      author = {Kentaro Tamura and Yutaka Shikano},
      title = {Quantum Random Number Generation with the Superconducting Quantum Computer IBM 20Q Tokyo},
      howpublished = {Cryptology ePrint Archive, Paper 2020/078},
      year = {2020},
      note = {\url{https://eprint.iacr.org/2020/078}},
      url = {https://eprint.iacr.org/2020/078}
}

@INPROCEEDINGS{9605294,

  author={Kuang, Randy and Lou, Dafu and He, Alex and McKenzie, Chris and Redding, Michael},

  booktitle={2021 IEEE International Conference on Quantum Computing and Engineering (QCE)}, 

  title={Pseudo Quantum Random Number Generator with Quantum Permutation Pad}, 

  year={2021},

  volume={},

  number={},

  pages={359-364},

  doi={10.1109/QCE52317.2021.00053}}

@ARTICLE{8396276,

  author={Truong, Nhan Duy and Haw, Jing Yan and Assad, Syed Muhamad and Lam, Ping Koy and Kavehei, Omid},

  journal={IEEE Transactions on Information Forensics and Security}, 

  title={Machine Learning Cryptanalysis of a Quantum Random Number Generator}, 

  year={2019},

  volume={14},

  number={2},

  pages={403-414},

  doi={10.1109/TIFS.2018.2850770}}

@misc{sinha2023programmable,
      title={A Programmable True Random Number Generator Using Commercial Quantum Computers}, 
      author={Aviraj Sinha and Elena R. Henderson and Jessie M. Henderson and Eric C. Larson and Mitchell A. Thornton},
      year={2023},
      eprint={2304.03830},
      archivePrefix={arXiv},
      primaryClass={quant-ph}
}

@misc{shi2022unbiased,
      title={An Unbiased Quantum Random Number Generator Based on Boson Sampling}, 
      author={Jinjing Shi and Tongge Zhao and Yizhi Wang and Chunlin Yu and Yuhu Lu and Ronghua Shi and Shichao Zhang and Junjie Wu},
      year={2022},
      eprint={2206.02292},
      archivePrefix={arXiv},
      primaryClass={quant-ph}
}

@misc{lanting2020probing,
      title={Probing Environmental Spin Polarization with Superconducting Flux Qubits}, 
      author={T. Lanting and M. H. Amin and C. Baron and M. Babcock and J. Boschee and S. Boixo and V. N. Smelyanskiy and M. Foygel and A. G. Petukhov},
      year={2020},
      eprint={2003.14244},
      archivePrefix={arXiv},
      primaryClass={quant-ph}
}

@inproceedings{krauss2021statistical,
  title={Statistical bias in D-wave qubits},
  author={Krauss, Thomas and Giffen, Alexander and Truppelli, Phillip and Michaels, Alan J},
  booktitle={Journal of Physics: Conference Series},
  volume={1936},
  number={1},
  pages={012010},
  year={2021},
  organization={IOP Publishing}
}

@article{turan2018recommendation,
  title={Recommendation for the entropy sources used for random bit generation},
  author={Turan, Meltem S{\"o}nmez and Barker, Elaine and Kelsey, John and McKay, Kerry A and Baish, Mary L and Boyle, Mike and others},
  journal={NIST Special Publication},
  volume={800},
  number={90B},
  pages={102},
  year={2018},
doi={10.6028/NIST.SP.800-90B},
url={https://nvlpubs.nist.gov/nistpubs/SpecialPublications/NIST.SP.800-90B.pdf}
}

@inproceedings{10.1145/3126908.3126925,
author = {Priedhorsky, Reid and Randles, Tim},
title = {Charliecloud: Unprivileged Containers for User-Defined Software Stacks in HPC},
year = {2017},
isbn = {9781450351140},
publisher = {Association for Computing Machinery},
address = {New York, NY, USA},
url = {https://doi.org/10.1145/3126908.3126925},
doi = {10.1145/3126908.3126925},
booktitle = {Proceedings of the International Conference for High Performance Computing, Networking, Storage and Analysis},
articleno = {36},
numpages = {10},
location = {Denver, Colorado},
series = {SC '17}
}

@ARTICLE{6773024,

  author={Shannon, C. E.},

  journal={The Bell System Technical Journal}, 

  title={A mathematical theory of communication}, 

  year={1948},

  volume={27},

  number={3},

  pages={379-423},

  doi={10.1002/j.1538-7305.1948.tb01338.x}}

@article{Baldwin_2022,
   title={Re-examining the quantum volume test: Ideal distributions, compiler optimizations, confidence intervals, and scalable resource estimations},
   volume={6},
   ISSN={2521-327X},
   url={http://dx.doi.org/10.22331/q-2022-05-09-707},
   DOI={10.22331/q-2022-05-09-707},
   journal={Quantum},
   publisher={Verein zur Forderung des Open Access Publizierens in den Quantenwissenschaften},
   author={Baldwin, Charles H. and Mayer, Karl and Brown, Natalie C. and Ryan-Anderson, Ciarán and Hayes, David},
   year={2022},
   month=may, pages={707} }

@article{PhysRevA.100.032328,
  title = {Validating quantum computers using randomized model circuits},
  author = {Cross, Andrew W. and Bishop, Lev S. and Sheldon, Sarah and Nation, Paul D. and Gambetta, Jay M.},
  journal = {Phys. Rev. A},
  volume = {100},
  issue = {3},
  pages = {032328},
  numpages = {11},
  year = {2019},
  month = {Sep},
  publisher = {American Physical Society},
  doi = {10.1103/PhysRevA.100.032328},
  url = {https://link.aps.org/doi/10.1103/PhysRevA.100.032328}
}

@article{Pelofske_2022,
   title={Quantum Volume in Practice: What Users Can Expect From NISQ Devices},
   volume={3},
   ISSN={2689-1808},
   url={http://dx.doi.org/10.1109/TQE.2022.3184764},
   DOI={10.1109/tqe.2022.3184764},
   journal={IEEE Transactions on Quantum Engineering},
   publisher={Institute of Electrical and Electronics Engineers (IEEE)},
   author={Pelofske, Elijah and Bärtschi, Andreas and Eidenbenz, Stephan},
   year={2022},
   pages={1–19} }

@inproceedings{matsuda2009quantum,
  title={Quantum annealing for problems with ground-state degeneracy},
  author={Matsuda, Yoshiki and Nishimori, Hidetoshi and Katzgraber, Helmut G},
  booktitle={Journal of Physics: Conference Series},
  volume={143},
  number={1},
  pages={012003},
  year={2009},
  organization={IOP Publishing},
  doi={10.1088/1742-6596/143/1/012003}
}

@article{PhysRevA.100.030303,
  title = {Uncertain fate of fair sampling in quantum annealing},
  author = {K\"onz, Mario S. and Mazzola, Guglielmo and Ochoa, Andrew J. and Katzgraber, Helmut G. and Troyer, Matthias},
  journal = {Phys. Rev. A},
  volume = {100},
  issue = {3},
  pages = {030303},
  numpages = {6},
  year = {2019},
  month = {Sep},
  publisher = {American Physical Society},
  doi = {10.1103/PhysRevA.100.030303},
  url = {https://link.aps.org/doi/10.1103/PhysRevA.100.030303}
}

@article{PhysRevLett.118.070502,
  title = {Exponentially Biased Ground-State Sampling of Quantum Annealing Machines with Transverse-Field Driving Hamiltonians},
  author = {Mandr\`a, Salvatore and Zhu, Zheng and Katzgraber, Helmut G.},
  journal = {Phys. Rev. Lett.},
  volume = {118},
  issue = {7},
  pages = {070502},
  numpages = {6},
  year = {2017},
  month = {Feb},
  publisher = {American Physical Society},
  doi = {10.1103/PhysRevLett.118.070502},
  url = {https://link.aps.org/doi/10.1103/PhysRevLett.118.070502}
}

@article{PhysRevE.99.063314,
  title = {Fair sampling of ground-state configurations of binary optimization problems},
  author = {Zhu, Zheng and Ochoa, Andrew J. and Katzgraber, Helmut G.},
  journal = {Phys. Rev. E},
  volume = {99},
  issue = {6},
  pages = {063314},
  numpages = {6},
  year = {2019},
  month = {Jun},
  publisher = {American Physical Society},
  doi = {10.1103/PhysRevE.99.063314},
  url = {https://link.aps.org/doi/10.1103/PhysRevE.99.063314}
}

@article{zhang2017advantages,
  title={Advantages of unfair quantum ground-state sampling},
  author={Zhang, Brian Hu and Wagenbreth, Gene and Martin-Mayor, Victor and Hen, Itay},
  journal={Scientific reports},
  volume={7},
  number={1},
  pages={1--12},
  year={2017},
  publisher={Nature publishing group},
  doi={10.1038/s41598-017-01096-6}
}

@article{marton2015interpretation,
  title={On the interpretation of results from the NIST statistical test suite},
  author={Marton, Kinga and Suciu, Alin},
  journal={Science and Technology},
  volume={18},
  number={1},
  pages={18--32},
  year={2015}
}

@Article{electronics12030723,
AUTHOR = {Crocetti, Luca and Nannipieri, Pietro and Di Matteo, Stefano and Fanucci, Luca and Saponara, Sergio},
TITLE = {Review of Methodologies and Metrics for Assessing the Quality of Random Number Generators},
JOURNAL = {Electronics},
VOLUME = {12},
YEAR = {2023},
NUMBER = {3},
ARTICLE-NUMBER = {723},
URL = {https://www.mdpi.com/2079-9292/12/3/723},
ISSN = {2079-9292},
DOI = {10.3390/electronics12030723}
}

@article{Guo:19,
author = {Xiaomin Guo and Chen Cheng and Mingchuan Wu and Qinzhong Gao and Pu Li and Yanqiang Guo},
journal = {Opt. Lett.},
number = {22},
pages = {5566--5569},
publisher = {Optica Publishing Group},
title = {Parallel real-time quantum random number generator},
volume = {44},
month = {Nov},
year = {2019},
url = {https://opg.optica.org/ol/abstract.cfm?URI=ol-44-22-5566},
doi = {10.1364/OL.44.005566},
}

@article{Bai_2021,
   title={18.8 Gbps real-time quantum random number generator with a photonic integrated chip},
   volume={118},
   ISSN={1077-3118},
   url={http://dx.doi.org/10.1063/5.0056027},
   DOI={10.1063/5.0056027},
   number={26},
   journal={Applied Physics Letters},
   publisher={AIP Publishing},
   author={Bai, Bing and Huang, Jianyao and Qiao, Guan-Ru and Nie, You-Qi and Tang, Weijie and Chu, Tao and Zhang, Jun and Pan, Jian-Wei},
   year={2021},
   month=jun }

@misc{D_Wave_postprocessing,
  title = {D-Wave Post Processing},
  howpublished = {\url{https://web.archive.org/web/20231122203833/https://docs.dwavesys.com/docs/latest/c_qpu_pp.html}}}

@dataset{pelofske_2024_10583977,
  author       = {Pelofske, Elijah},
  title        = {{Dataset for Analysis of a Programmable Quantum 
                   Annealer as a Random Number Generator}},
  month        = jan,
  year         = 2024,
  publisher    = {Zenodo},
  doi          = {10.5281/zenodo.10583977},
  url          = {https://doi.org/10.5281/zenodo.10583977}
}

@article{santoro2006optimization,
  title={Optimization using quantum mechanics: quantum annealing through adiabatic evolution},
  author={Santoro, Giuseppe E and Tosatti, Erio},
  journal={Journal of Physics A: Mathematical and General},
  volume={39},
  number={36},
  pages={R393},
  year={2006},
  publisher={IOP Publishing},
doi={10.1088/0305-4470/39/36/R01}
}

@article{Finnila_1994,
   title={Quantum annealing: A new method for minimizing multidimensional functions},
   volume={219},
   ISSN={0009-2614},
   url={http://dx.doi.org/10.1016/0009-2614(94)00117-0},
   DOI={10.1016/0009-2614(94)00117-0},
   number={5–6},
   journal={Chemical Physics Letters},
   publisher={Elsevier BV},
   author={Finnila, A.B. and Gomez, M.A. and Sebenik, C. and Stenson, C. and Doll, J.D.},
   year={1994},
   month=mar, pages={343–348} }

\end{document}